\documentclass{amsart}
\usepackage{amssymb,amsmath}
\usepackage{natbib}
\usepackage[ruled, norelsize]{algorithm2e}
\usepackage[left=1.5in,right=1.5in,top=1in,bottom=1in]{geometry}
\usepackage[pdftex]{graphicx}

\def\iid{\ensuremath{\stackrel{\text{\tiny iid}}{\sim}}}

\def\argmax{\mathop{\rm arg\,max}\limits}%
\let\hat\widehat
\def\given{{\,|\,}}
\def\Pr{{\ensuremath{\mathbb P}}}%
\def\Exp{{\ensuremath{\mathbb E}}}%
\def\eop{{\textbf{e}}}%
\def\V{{\ensuremath{\mbox{\rm vec}}}}%
\def\tr{{\ensuremath{\mbox{\rm tr}}}}%
\def\cov{{\ensuremath{\mbox{\rm Cov}}}}%
\def\var{{\ensuremath{\mbox{\rm Var}}}}%

\begin{document}
\title{An Expectation-Maximization Algorithm for the Matrix Normal
Distribution}

\author{Hunter Glanz \and Luis Carvalho}
\thanks{
H. Glanz and L. Carvalho
at Dept. of Mathematics and Statistics, Boston University, Boston, MA, 02215 USA}

\begin{abstract}
Dramatic increases in the size and dimensionality of many recent data sets
make crucial the need for sophisticated methods that can exploit inherent
structure and handle missing values. In this article we derive an
expectation-maximization (EM) algorithm for the matrix normal distribution, a
distribution well-suited for naturally structured data such as spatio-temporal
data. We review previously established maximum likelihood matrix normal
estimates, and then consider the situation involving missing data. We apply
our EM method in a simulation study exploring errors across different
dimensions and proportions of missing data. We compare these errors and
computational running times to those from two alternative methods. Finally, we
implement the proposed EM method on a satellite image dataset to investigate
land-cover classification separability.
\end{abstract}


\maketitle

\section{Introduction}
Technological advances in recent decades have ushered in a new era in data
management and analysis. The dimension of data sets continues to grow
alongside the number of observations. Consequently, the estimation of
parameters or characteristics of these data remains a significant challenge.
Specifically, the covariance matrix of such high dimensional data can be
extremely difficult to estimate and handle. An increasingly common
simplification is the assumption that this covariance has a kronecker product
structure.

Not only does this structure ease the estimation procedure, it naturally fits
many situations in a more physical way. Multivariate repeated measures data,
for example, provides this type of framework
\citep{boik,naik,roykhat05,roy05}. That is, the response variables and time are
two separate \emph{dimensions} of the data that can be characterized
independently. Thus, a straightforward way to model the full covariance is via
the kronecker product of a covariance of the responses and a covariance of
time. 

Similarly, multivariate time series of other types such as longitudinal data
\citep{chaganty,galecki} and spatio-temporal data \citep{shitan,fuentes} lend
themselves to this kind of covariance decomposition. To this end, classic
work has been done on estimates for the matrix normal distribution:
\begin{equation}
\label{eq:matlik}
X \sim MN(\mu, \Sigma_s, \Sigma_c),
\end{equation}
where $X$ and $\mu$ are $p \times q$ matrices, $\Sigma_s$ is a $p \times p$
matrix and $\Sigma_c$ is a $q \times q$ matrix
\citep{dawid,pierre,sriv79,sriv07}. The matrix normal distribution is
synonymous with this kronecker covariance structure since if $X$ has the above
matrix normal distribution, then
\[
	\V(X) \sim N(\V(\mu), \Sigma_c \otimes \Sigma_s),
\]
where $\Sigma_c \otimes \Sigma_s$ denotes the kronecker product of $\Sigma_c$
and $\Sigma_s$ and $\V(X_i)$ denotes the vectorization of $X_i$. While maximum
likelihood estimates have been derived along with tests of whether a
covariance matrix has this structure, the issue of missing data in this
context has not been addressed. In this article we derive an
expectation-maximization (EM) algorithm~\citep{em} for estimating the
parameters of the matrix normal distribution when missing data exists. With
expressions for the estimates in hand, we conduct a simulation study and apply
the method to remotely sensed data collected from the Moderate Resolution
Imaging Spectroradiometer (MODIS) sensor aboard the AQUA and TERRA satellite
platforms~\citep{friedl}.

\section{Methods}
We begin with $N$ independent observations from a matrix normal distribution,
\[
X_i \iid MN(\mu, \Sigma_s, \Sigma_c), \quad i = 1, \dots, N,
\]
where $X_i$ and $\mu$ are $p \times q$ matrices, $\Sigma_s$ is $p \times p$
and $\Sigma_c$ is $q \times q$. The use of this structure reduces the number
of parameters by explicitly describing the covariance between the rows and the
covariance between the columns as opposed to an individual covariance in each
cell of the upper triangle of the full, $pq \times pq$, covariance matrix.
Besides this simplification, the partitioning of the covariance follows
naturally from a setup involving two \emph{physical}, or \emph{separable},
dimensions such as space and time.

\subsection{Parameter Estimation with Complete Data}
It is straightforward to estimate the parameters of a matrix normal
distribution using maximum likelihood when there is no missing data.
If $X_i \iid MN(\mu, \Sigma_s, \Sigma_c)$ then, equivalently,
\[
\V(X_i) \iid N(\V(\mu), \Sigma_c \otimes \Sigma_s)
\]
and so
\begin{multline*}
\Pr(X_i; \mu, \Sigma_c, \Sigma_s) = 
(2\pi)^{-\frac{pq}{2}} |(\Sigma_c \otimes \Sigma_s)^{-1}|^{\frac{1}{2}} \\
\exp\Bigg\{ -\frac{1}{2}
\V(X_i-\mu)^\top(\Sigma_c \otimes \Sigma_s)^{-1}\V(X_i-\mu) \Bigg\}.
\end{multline*}

As usual in multivariate normal densities, in the exponential term we have the
Mahalanobis distance
$D_{\Sigma}(X_i, \mu)$, with $\Sigma := \Sigma_c \otimes \Sigma_s$, between
$X_i$ and $\mu$,
\[
D_{\Sigma}(X_i, \mu) \doteq 
\V(X_i-\mu)^\top (\Sigma_c \otimes \Sigma_s)^{-1}\V(X_i-\mu).
\]
This distance can be worked through known identities of the vec operator and
kronecker product to yield a simpler expression for the matrix normal density:
\[
\begin{split}
D_{\Sigma}(X_i, \mu) &= 
\V(X_i-\mu)^\top(\Sigma_c \otimes \Sigma_s)^{-1}\V(X_i-\mu) \\
&\stackrel{\text{\tiny (i)}}{=} 
\V(X_i-\mu)^\top(\Sigma_c^{-1} \otimes \Sigma_s^{-1})\V(X_i-\mu) \\
&\stackrel{\text{\tiny (ii)}}{=} 
\V(X_i-\mu)^\top \V[\Sigma_s^{-1} (X_i - \mu) \Sigma_c^{-1}] \\
&\stackrel{\text{\tiny (iii)}}{=} 
\tr \Big[(X_i-\mu)^\top \Sigma_s^{-1} (X_i - \mu) \Sigma_c^{-1} \Big] \\
&=
\tr \Big[\Sigma_c^{-1} (X_i-\mu) \Sigma_s^{-1} (X_i - \mu)^\top \Big]
\end{split}
\]
where (i), (ii), and (iii) are applications of identities (488), (496), and
(497) in \citep{petersen08}, respectively. Furthermore, since
\[
|(\Sigma_c \otimes \Sigma_s)^{-1}| =
|\Sigma_c^{-1}|^p |\Sigma_s^{-1}|^q
\]
by identity (492) in \citep{petersen08}, we recover the characterization of
\citet{sriv79}: $X_i \sim MN(\mu, \Sigma_s, \Sigma_c)$
if and only if the density of $X_i$ is given by
\begin{multline}
\label{eq:nicelik}
\Pr(X_i; \mu, \Sigma_c, \Sigma_s) = 
(2\pi)^{-\frac{pq}{2}} |\Sigma_s|^{-\frac{q}{2}} |\Sigma_c|^{-\frac{p}{2}} \\
\exp\Bigg\{ -\frac{1}{2} \tr\Big\{
\Sigma_c^{-1}(X_i - \mu)\Sigma_s^{-1}(X_i - \mu)^\top \Big\} \Bigg\}.
\end{multline}

The log likelihood of the parameters $\Theta = (\mu, \Sigma_s, \Sigma_c)$
is then
\begin{multline}
\label{eq:full.lik}
\log \Pr(X_1, \ldots, X_n; \Theta) = \sum_i \log \Pr(X_i; \Theta) \\
\qquad = \frac{pN}{2} \log|\Sigma_c^{-1}| + \frac{qN}{2} \log|\Sigma_s^{-1}|
-\frac{N}{2} \log\sigma^2 \\
- \frac{1}{2\sigma^2} \sum_i D_{\Sigma}(X_i, \mu)
\end{multline}
up to a normalizing constant. The form in~\eqref{eq:nicelik} simplifies the
matrix derivatives of~\eqref{eq:full.lik} considerably leaving us with the
following maximum likelihood estimates (MLEs):
\begin{equation}
\label{eq:mle}
\begin{split}
\hat{\mu} &= \frac{1}{N} \sum_{i=1}^N X_i \\
\hat{\Sigma}_c &= \frac{1}{pN}
\sum_{i=1}^N (X_i - \hat{\mu})^\top \hat{\Sigma}_s^{-1} (X_i - \hat{\mu}) \\
\hat{\Sigma}_s &= \frac{1}{qN}
\sum_{i=1}^N (X_i - \hat{\mu}) \hat{\Sigma}_c^{-1} (X_i - \hat{\mu})^\top
\end{split}
\end{equation}
The two covariance estimates depend on each other and thus their estimates
must be computed in an iterative fashion until convergence.

\subsubsection*{Handling non-identifiability}
If for any $\kappa \neq 0$ we define $\tilde{\Sigma}_c := \kappa \Sigma_c$ and
$\tilde{\Sigma}_s := \Sigma_s / \kappa$ then
$\tilde{\Sigma}_c \otimes \tilde{\Sigma}_s
= \Sigma_c \otimes \Sigma_s$,
and so both estimates yield the same overall covariance matrix. To resolve
this non-identifiability issue we propose the following amendment to the
model:
\begin{equation}
\label{eq:convlik}
\V(X_i) \sim N(\V(\mu), \sigma^2 \Sigma_c \otimes \Sigma_s)
\end{equation}
and require that $(\Sigma_c)_{11} = 1$ and $(\Sigma_s)_{11} = 1$ (the choice
of the top-left entry is arbitrary.) In this way we fix the scale of
$\Sigma_c$ and $\Sigma_s$, and estimate the scale of the overall covariance in
$\sigma^2$.

The MLE of $\sigma^2$ is
\[
\hat{\sigma}^2 = \frac{1}{pqN} \sum_{i=1}^N
(X_i - \hat{\mu})^\top (\hat{\Sigma}_c \otimes \hat{\Sigma}_s)^{-1}
(X_i - \hat{\mu})
\]
and depends on the other estimates. The MLE for $\mu$ is clearly the same as
in~\eqref{eq:mle} since we only changed the variance of the model. However,
since the variance scale is now captured by $\sigma^2$ we need to scale the
MLEs for $\Sigma_c$ and $\Sigma_s$ by their top-left entry at each
iteration: if $\hat{\Sigma}^*_c$ and $\hat{\Sigma}^*_s$ are the estimates
from~\eqref{eq:mle} for $\Sigma_c$ and $\Sigma_s$, then the respective MLEs
for~\eqref{eq:convlik} are
$\hat{\Sigma}_c = \hat{\Sigma}^*_c / (\hat{\Sigma}^*_c)_{11}$ and
$\hat{\Sigma}_s = \hat{\Sigma}^*_s / (\hat{\Sigma}^*_s)_{11}$.

Finally, we remark that, according to \citet[Theorem~3.1]{sriv07}, if $N >
\max(p,q)$ then the maximum likelihood estimates are unique.

\subsection{Parameter Estimation with Missing Data}
\label{sec:em}
Missing data presents a difficult, albeit well-studied challenge in parameter
estimation. Traditional methods, such as the EM algorithm, can usually handle
missing data in a straightforward way. As the dimensionality increases, as in
our case, the method can become quite computationally expensive.
Naturally, we aim to assess different ways of achieving accurate parameter
estimates with an eye towards reducing computation time. 

The first approach (which we label ``$MM$'') applies a maximization in two
ways: 1) ``imputation" of missing values and 2) maximum likelihood parameter
estimation. In particular, the missing values get replaced by the most recent
estimate of the mean. The next iteration of mean and covariance estimates come
from the same maximum likelihood expressions in~\eqref{eq:mle}, with the
addition of $\hat{\sigma}^2$, based on the fully imputed data. The ease and
simplicity of this method make it a natural first step in handling missing
data, but also hinder its robustness and ability to capture all of the
uncertainty associated with missing data.

The second approach (``$GEM$'') applies the EM algorithm to the most general
version of the model. As opposed to estimating the parameters
of~\eqref{eq:convlik}, the EM algorithm provides parameter estimates for the
following model:
\begin{equation}
\label{eq:gemlik}
\V(X_i) \sim N(\V(\mu), \Sigma).
\end{equation}
These multivariate normal EM estimates have the same form of those found in
\cite{gembook}. This approach does not simplify the original problem since it
requires more parameters to be estimated by not assuming the kronecker
structure. The simple form of~\eqref{eq:gemlik} attracts much attention, but
its complexity far exceeds that of~\eqref{eq:convlik}. Where sources of
variation in the data can be naturally partitioned, such as in space or time,
the kronecker structure surpasses~\eqref{eq:gemlik} in both interpretability
and computational efficiency. The remainder of this article explains an EM
procedure for~\eqref{eq:convlik} and its superiority in situations involving
these types of structured data.

\subsubsection*{EM Algorithm for Matrix Normal Distribution}
In the situation where missing data exists the EM algorithm is a
convenient way to estimate the parameters $\Theta = (\mu, \Sigma_c, \Sigma_s,
\sigma^2)$ in~\eqref{eq:convlik}. The rest of this section details the third
approach (``$EM$'') to parameter estimation with missing data.
Let us denote $X=(Y,Z)$ where $Z$ is the missing portion of $X$ and $Y$ is the
observed portion of $X$.
For the \emph{E-step}, we need:
\begin{multline*}
Q(\Theta; \Theta^{(t)}) \doteq
\Exp_{Z \given Y; \Theta^{(t)}} [\log \Pr(X_1, \ldots, X_n; \Theta)] \\
\qquad = \frac{pN}{2} \log|\Sigma_c^{-1}| + \frac{qN}{2} \log|\Sigma_s^{-1}|
-\frac{N}{2} \log\sigma^2 \\
- \frac{1}{2\sigma^2} \sum_i
\Exp_{Z \given Y; \Theta^{(t)}} [D_{\Sigma}(X_i, \mu)],
\end{multline*}
while the \emph{M-step} updates $\Theta$ by maximizing $Q$,
\[
\Theta^{(t+1)} \doteq \argmax_{\Theta} Q(\Theta; \Theta^{(t)}),
\]
via matrix differentiation in our case.

The Mahalanobis distance obeys a Pythagorean relationship: if
$\tilde{\mu}_i^{(t)} := \Exp_{Z \given Y; \Theta^{(t)}}[X_i]$, then
\[
\Exp_{Z \given Y; \Theta^{(t)}} [D_{\Sigma}(X_i, \mu)] = 
 \Exp_{Z \given Y; \Theta^{(t)}} [D_{\Sigma}(X_i, \mu_i^{(t)})] +
D_{\Sigma}(\mu_i^{(t)}, \mu).
\]
From here the update for $\mu$ follows from
$\partial Q/\partial \mu = 0$:
\[
\hat{\mu}^{(t+1)} = \frac{1}{N} \sum_{i=1}^N \tilde{\mu}_i^{(t)},
\]
similarly to the plain MLE case in~\eqref{eq:mle}.

Updating $\Sigma_c$, $\Sigma_s$ and $\sigma^2$ requires a bit more work. To
this end we focus, first, on the following term:
\begin{multline*}
R_i(\Theta; \Theta^{(t)}) \doteq 
\Exp_{Z \given Y; \Theta^{(t)}} [D_{\Sigma}(X_i, \mu_i^{(t)})] \\
= \Exp\Big[\tr[(\Sigma_c \otimes
\Sigma_s)^{-1}\V(X_i-\tilde{\mu}_i^{(t)})\V(X_i-\tilde{\mu}_i^{(t)})^\top]\Big] \\
= \tr\Big[(\Sigma^{-1}_c \otimes \Sigma^{-1}_s)
\underbrace{%
\Exp[\V(X_i-\tilde{\mu}^{(t)}_i)\V(X_i-\tilde{\mu}^{(t)}_i)^\top]}_{V_i^{(t)}}
\Big].
\end{multline*}
where we define the expected outer product
\[
V_i^{(t)} \doteq \Exp_{Z \given Y; \Theta^{(t)}}
[\V(X_i - \tilde{\mu}^{(t)}_i)\V(X_i - \tilde{\mu}^{(t)}_i)^\top].
\]

To get the partial derivatives of $Q$ with respect to $\Sigma_c$ we need
\[
\begin{split}
\frac{\partial R_i}{\partial (\Sigma_c^{-1})_{kl}} & = \tr\left\{ \frac{\partial}{\partial (\Sigma_c^{-1})_{kl}} [(\Sigma_c^{-1} \otimes \Sigma_s^{-1})V_i] \right\} \\
& = \tr \left\{ \left( 
\underbrace{%
\frac{\partial\Sigma_c^{-1}}{\partial (\Sigma_c^{-1})_{kl}}}_{S_{kl}} \otimes \Sigma_s^{-1} \right)V_i \right\} 
\end{split}
\]
where $S_{kl}$ is the structure matrix \citep{petersen08} of a symmetric
matrix, that is,
$S_{kl} = [\delta_{ik} \cdot \delta_{jl} + \delta_{il} \cdot
\delta_{jk}]_{ij}$.
Thus, $(S_{kl} \otimes \Sigma_s^{-1})$ is a block matrix.

We can now look at $V_i$ as a $q \times q$ block matrix where each element is
a $p \times p$ matrix in the following way, for example:
\begin{equation}
\label{eq:bandmatrix}
\hspace{.8in}
\begin{matrix}
& b & \\
V_{i, kl} = &
\begin{bmatrix}
\circ & \circ   & \circ & \circ   & \circ & \circ   & \circ \\
\circ & \bullet & \circ & \bullet & \circ & \bullet & \circ \\
\circ & \circ   & \circ & \circ   & \circ & \circ   & \circ \\
\circ & \bullet & \circ & \bullet & \circ & \bullet & \circ \\
\circ & \circ   & \circ & \circ   & \circ & \circ   & \circ
\end{bmatrix}
&
\begin{matrix} \\ b' \\ \\ \\ \\ \end{matrix}
\end{matrix}
\end{equation}
%
So the matrix $V_i$ being a block matrix leads to $V_{i, kl}$ being a
symmetric $p \times p$ matrix with zeros at the empty circles
in~\eqref{eq:bandmatrix} and $\cov_{Z_i \given Y_i}(Z_{kb}, Z_{lb'})$ at the
filled circles. Thus,
\begin{multline*}
\frac{\partial R_i}{\partial (\Sigma_c^{-1})_{kl}} =
\tr[\Sigma_s^{-1} (V_{i, kl} + V_{i, kl}^\top)] \\
= \sum_{b, b' \in \text{\sf miss}(k, l)} (\Sigma_s^{-1})_{b, b'}
\cov_{Z_i \given Y_i, \Theta^{(t)}} [Z_{kb}, Z_{lb'}].
\end{multline*}
Here {\sf miss}$(k, l)$ are the row-column pairs for which there are missing
entries in $V_i$, and $V_{i, kl}$ is the $p \times p$ block submatrix of $V_i$ from
rows $(k-1)p + 1$ to $kp$ and columns $(l-1)p + 1$ to $lp$. Note that
$\partial R_i/\partial (\Sigma_c^{-1})_{k,l}$ does not depend on
$Z_i$. Moreover, the conditional covariances in
$\var_{Z_i \given Y_i, \Theta^{(t)}}[Z_i]$ above can be obtained by applying
the SWEEP operator~\citep{goodnight79} to the rows of
$\Sigma_c^{-1} \otimes \Sigma_s^{-1}$ that correspond to missing values.

Thus, from solving $\partial Q/\partial (\Sigma_c^{-1}) = 0$, we have
%
%
%
\begin{equation}
\label{eq:sigcup}
\hat{\Sigma}_c^{(t+1)} = \frac{1}{\hat{\sigma}^{2^{(t)}}pN} \sum_i \Bigg(
\frac{\partial R_i}{\partial \Sigma_c^{-1}} 
+ (\tilde{\mu}^{(t)}_i - \mu^{(t)})^\top \Sigma_s^{-1^{(t)}}
(\tilde{\mu}^{(t)}_i - \mu^{(t)}) \Bigg).
\end{equation}
Similarly, for $\hat{\Sigma}_s$,
\begin{equation}
\label{eq:sigsup}
\hat{\Sigma}_s^{(t+1)} = \frac{1}{\hat{\sigma}^{2^{(t)}}qN} \sum_i \Bigg(
\frac{\partial R_i}{\partial \Sigma_s^{-1}} \\
+ (\tilde{\mu}^{(t)}_i - \mu^{(t)}) \Sigma_c^{-1^{(t)}}
(\tilde{\mu}^{(t)}_i - \mu^{(t)})^\top \Bigg).
\end{equation}
Finally, for $\sigma^2$,
\begin{equation}
\label{eq:sig2up}
\hat{\sigma}^{2^{(t+1)}} = \frac{1}{pqN} \sum_i R_i \\
+ \V(\tilde{\mu}^{(t)}_i-\hat{\mu}^{(t)})^\top(\hat{\Sigma}^{(t)}_c \otimes
\hat{\Sigma}^{(t)}_s)^{-1}\V(\tilde{\mu}^{(t)}_i-\hat{\mu}^{(t)}).
\end{equation}
Just as in the last section, we normalize these covariance estimates by their
upper-left entry; i.e. $(\hat{\Sigma}_c)_{11}$ and $(\hat{\Sigma}_s)_{11}$.
Detailed information regarding the implementation of this EM method can be
found in the Appendix, with \verb!R! code included in the supplementary material.

\section{Case Studies}
\subsection{Simulation Study}
To empirically assess the model and algorithm we simulated data from a matrix
normal distribution with randomly chosen parameters of dimensions: $p = 3$ and
$q = 5$, and $p = 3$ and $q = 7$.
Three sample sizes were used: 250, 500, and 1000. Four different proportions
of missing data were used: 5\%, 10\%, 15\% and 20\%. Data was simulated 100
times at each combination of sample size and proportion of missing data to
evaluate how the accuracy of the estimates vary. The three different
algorithms described in Section~\ref{sec:em} were run in each of these
combinations to provide a richer comparison. In order to compare these
methods, the covariance errors were always measured with respect to the full
(kronecker product) covariance matrix.


\begin{figure*}
	\centerline{
		\includegraphics[width=6in]{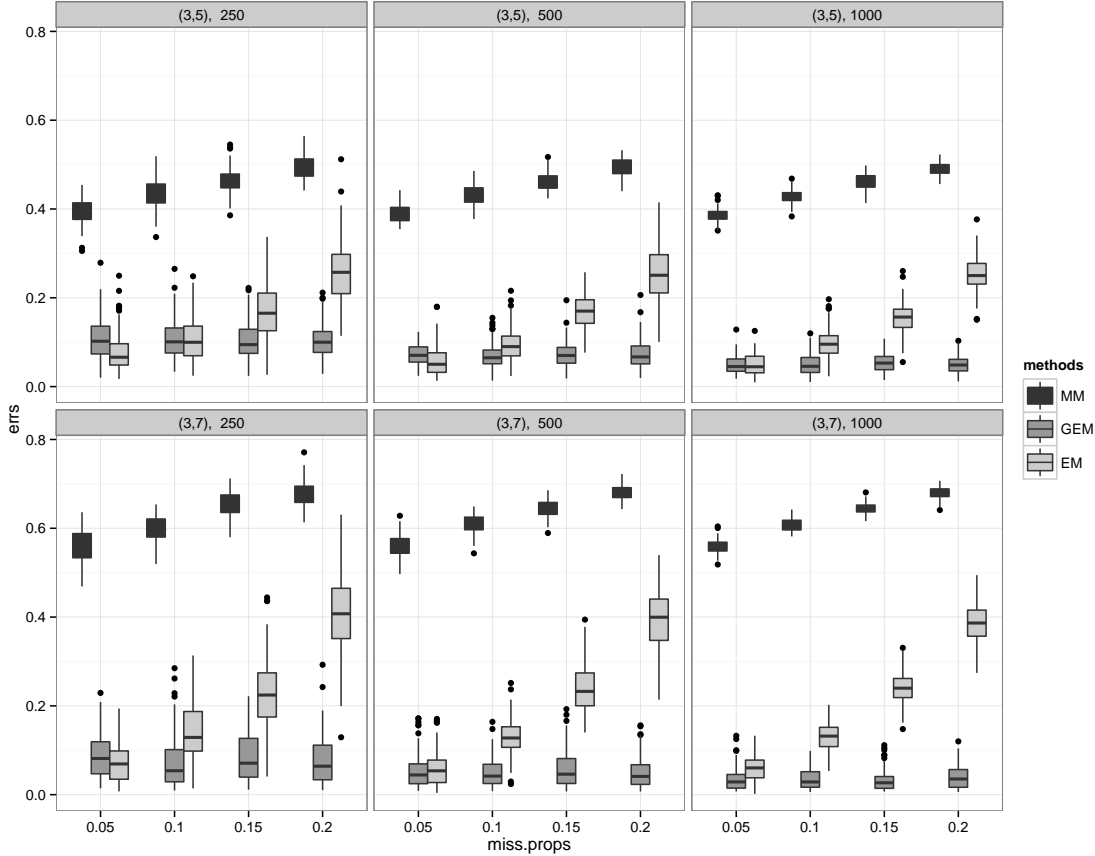}
	}
	\caption{Boxplots of the relative errors in the estimates of $\Sigma$, in order of $MM$, $GEM$, and $EM$.}
\label{fig:sigboxp}
\end{figure*}

\begin{figure*}
	\centerline{
		\includegraphics[width=6in]{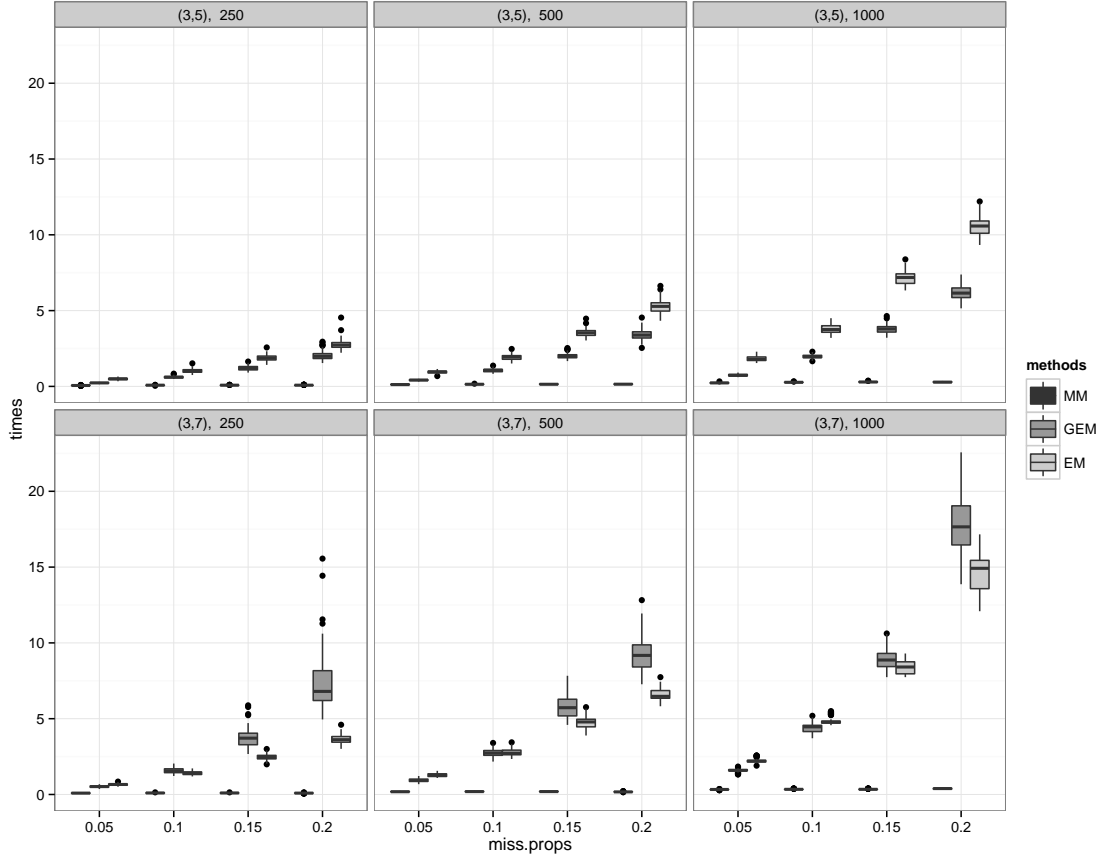}
	}
	\caption{Boxplots of the run times (in seconds), in order of $MM$, $GEM$, and $EM$.}
\label{fig:timeboxp}
\end{figure*}

The relative errors of the mean estimates across the three methods and the
four different proportions of missing data were consistently low. The methods
differ very little when it comes to the estimate of the mean, and so we focus
on the variance estimates. Indeed, the models and estimation procedure vary
most when dealing with the covariance matrix.

Figure~\ref{fig:sigboxp} tells a rich story about how these three methods
differ most. Since the $MM$ method essentially treats the imputed values as
actual data and fails to account for all of the uncertainty present, the
errors for this method top all of those from the $GEM$ and $EM$. As the sample
size increases the estimates appear to improve slightly, but the most
interesting feature lies in the difference between the $GEM$ and $EM$ methods.
Since~\eqref{eq:gemlik} contains more parameters to be estimated, this model
can achieve better resolution and accuracy than~\eqref{eq:convlik}, albeit
needing more samples to identify parameters. The significant cost lies in the
computation time, making the kronecker structure a worthwhile consideration
since it noticeably reduces the complexity of the model and still achieves
accurate parameter estimates.

Figure~\ref{fig:timeboxp} introduces the computational differences between the
methods. The $MM$ method, while still requiring some iterating, takes very
little time in all scenarios. Interestingly, the $EM$ method requires the most
time for lower dimensions such as the $p = 3$, $q = 5$ situation simulated
here. When the dimensions increase, we begin to see the gains in the kronecker
model. Naturally, as the discrepancy in the number of parameters being
estimated by $GEM$ and $EM$ increases, the computational advantages become
more significant. Additionally, the difference in sample sizes necessary to
estimate the covariance grows, with $GEM$ requiring more.

Of course, this presumes the choice between the two models. In a situation
where the physical dimensions of the data imply a kronecker structure, we can
take comfort in the above results. One such example is the following
application to Remote Sensing.


\subsection{Land Cover Classification using {MODIS} Satellite Image Data}
One of the biggest tasks in Remote Sensing is land-cover classification. In
other words, taking remotely sensed images composed of millions of pixels and
assigning to each pixel a particular land cover class. Many of the richest
datasets are both multispectral and multitemporal. For satellite image data
from the MODIS device we observe data in 7 different spectral bands at each of
46 time points (comprising a multivariate annual profile). The analysis we
present here is an application of the proposed EM algorithm to these
multivariate time series data. We adopt the following model:
\begin{equation}
\label{eq:geolik}
\V(X_v) \given \theta_v = c \sim N(\V(\mu_c), \sigma^2_c \Sigma_c \otimes \Sigma_s),
\end{equation}
where $X_v$ and $\theta_v$ are the data and land cover class for pixel $v$,
respectively. In this geographical context the kronecker structure naturally
models the covariance in the spectral bands ($\Sigma_s$) and the covariance in
time for each class ($\Sigma_c$). That is, the full covariance is partitioned
into spectral and temporal components. Note that in (\ref{eq:geolik}) we allow
for a different mean, temporal covariance and scale ($\sigma^2$) for each
class. 

For geographical reasons we focus our attention on the middle of the year (the
middle 28 time points). Thus $X$ and $\mu_c$ are $7 \times 28$ matrices,
$\Sigma_c$ is a $28 \times 28$ matrix and $\Sigma_s$ is a $7 \times 7$ matrix.
For the analysis that follows we used a subset of the MODIS Land Cover
Training site database that includes 204 sites located over the conterminous
United States. These sites include 2,245 MODIS pixels and encompass most major
biomes and land cover types in the lower 48 United States~\citep{schaaf}. 

Land cover classification has been approached in many different ways for quite
some time. The high dimensionality of the data presents an increasingly
significant challenge. Traditionally, principal components analysis (PCA) has
been used to reduce the dimensionality of the data. However, usually all 196
($7 \cdot 28$) features are considered distinct features. Consequently,
spectral and temporal variation cannot be isolated from an
eigen-decomposition of the full ($196 \times 196$) covariance matrix, as
required from PCA. 

To preserve the raw temporal information and still reduce the dimensionality
of the data, we propose targeting $\Sigma_s$ with the principal components
analysis as opposed to $\Sigma_c \otimes \Sigma_s$. About 5\% of the data is
missing, making the kronecker structure and our proposed EM algorithm even
more ideal, from a purely computational standpoint, as evidenced by the
simulation study above. Using the proposed EM algorithm we estimate the
parameters of~\eqref{eq:geolik} and perform PCA on our estimate of $\Sigma_s$.
To, again, provide a comparison which confirms the quality of our $EM$ we
compare the PCA results just described to those computed using the $MM$
method.

\begin{figure*}
\centerline{
  \includegraphics[width=3.5in]{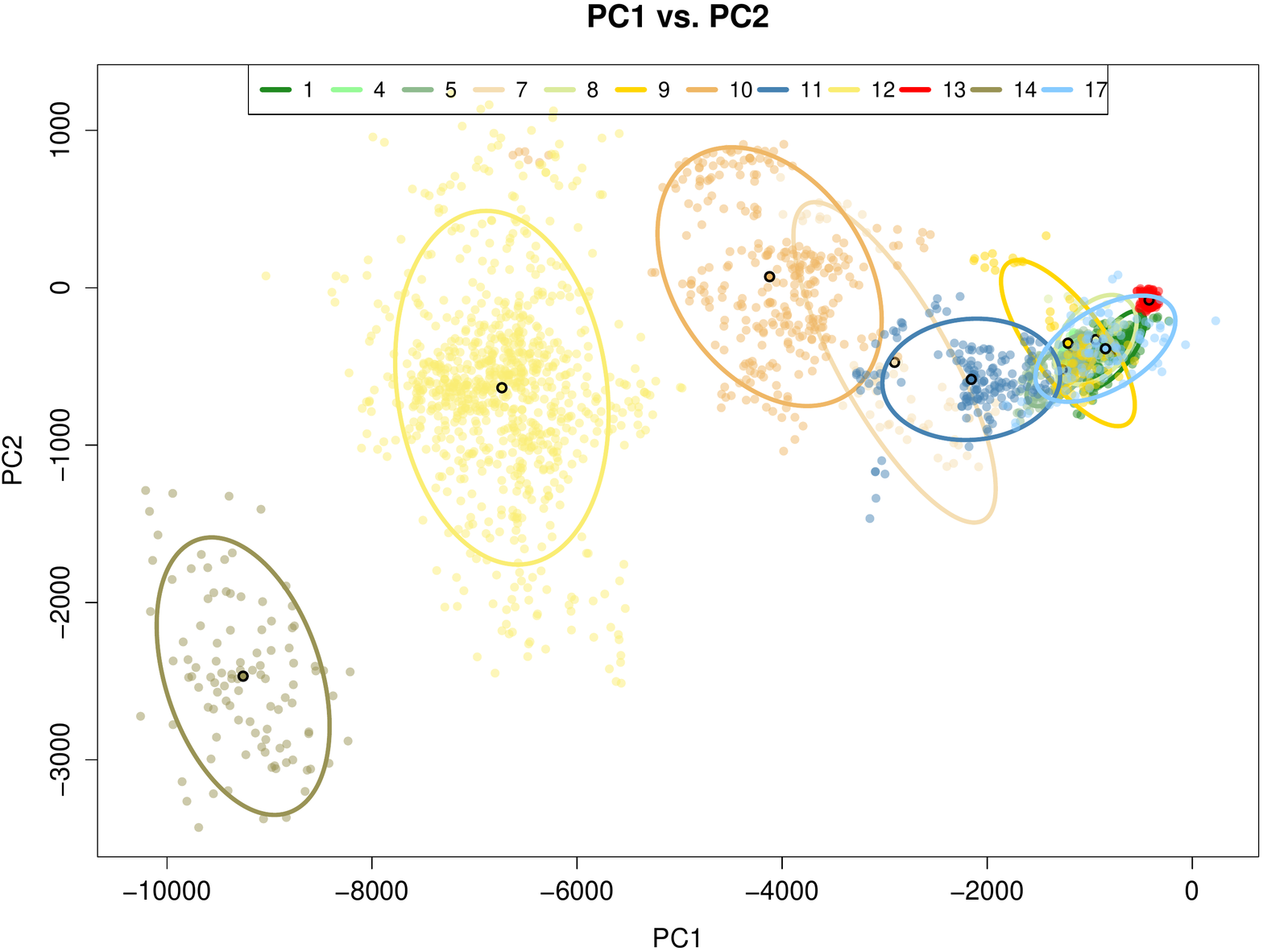}
  \includegraphics[width=3.5in]{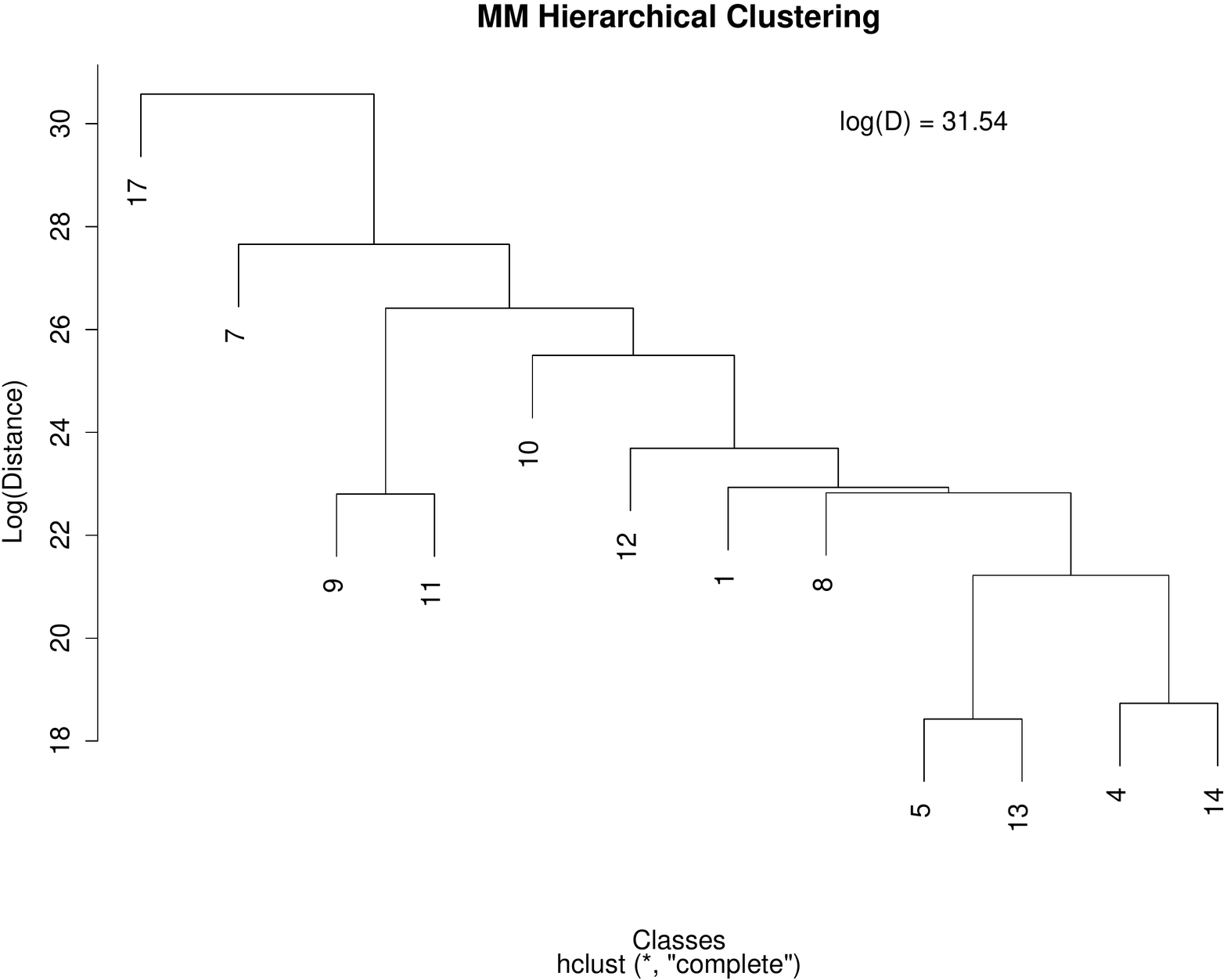}
}
\centerline{
  \includegraphics[width=3.5in]{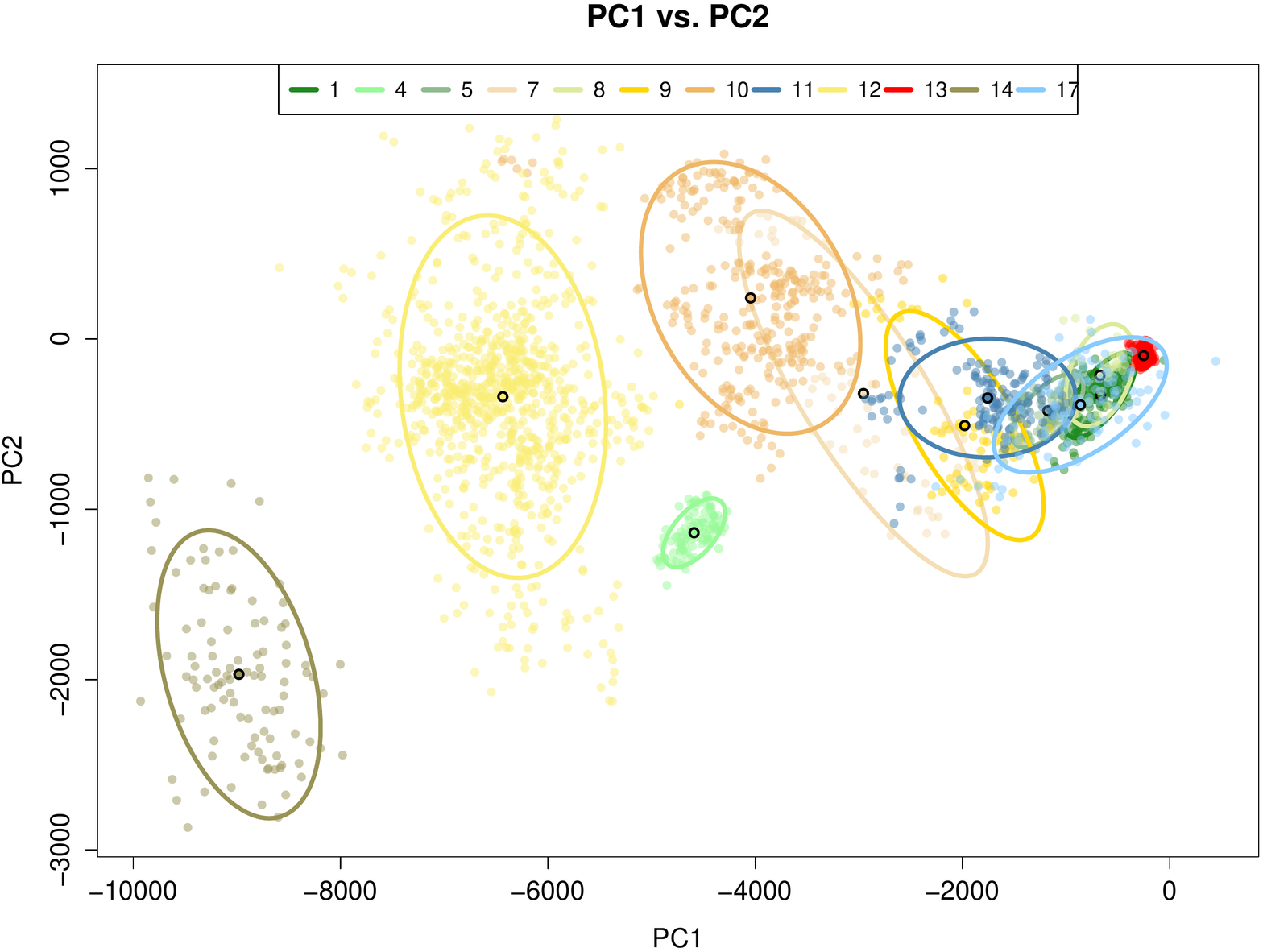}
  \includegraphics[width=3.5in]{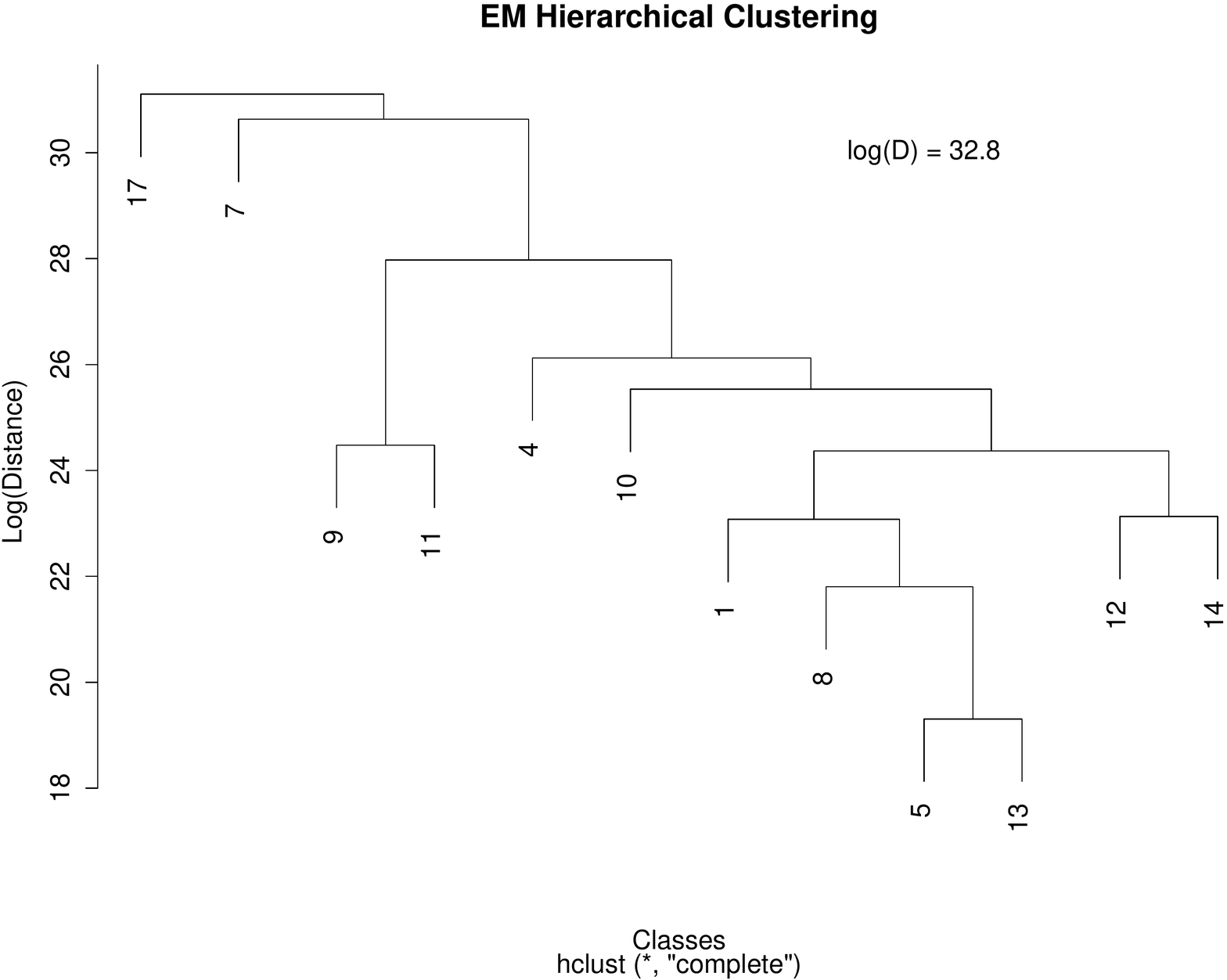}
}
\caption{Left: the data projected into the space of the first two principal
components, colored by their respective land color classes; right:
hierarchical clustering of classes. Top and bottom panels: results using MM
and EM, respectively.}
\label{fig:geopc}
\end{figure*}

Figure~\ref{fig:geopc} shows the data projected into the space of the first
two principal components using the $MM$ and $EM$ methods. The class
separability seems reasonable and the different amounts of variation in each
class are certainly identified. A closer inspection reveals that the $EM$
method achieves better separability of the land-cover classes. This can be
seen in the PC1 versus PC2 plots as well as their neighboring dendrograms.
These dendrograms resulted from a hierarchical clustering of the classes using
the following distance metric, $d_{ij}$:
\[
d_{ij} = (\mu_i - \bar{\mu}_{ij})^\top \Sigma_i^{-1}(\mu_i - \bar{\mu}_{ij})
+ (\mu_j - \bar{\mu}_{ij})^\top \Sigma_j^{-1}(\mu_j - \bar{\mu}_{ij}).
\]
where the $\bar{\mu}_{ij}$ are ``in-between'' centers for clusters $i$ and
$j$,
\[
\bar{\mu}_{ij} = \Sigma_i^{-1}(\Sigma_i^{-1} + \Sigma_j^{-1})^{-1}\mu_i 
+ \Sigma_j^{-1}(\Sigma_i^{-1} + \Sigma_j^{-1})^{-1}\mu_j.
\]
The log distances are plotted above along with an overall log distance,
$D = \sum_{i < j} d_{ij}$.
As a measure of the overall separability of the classes in the space of the
first two principal components, the greater value of $\log(D)$ achieved by the
$EM$ confirms that this method helped to better distinguish land-cover
classes.

In particular, using the $EM$, the first two principal components capture
76.1\% of the variation in the spectral bands. For the purposes of land-cover
classification we chose to use the first \emph{three} principal components
since they capture 92.3\% of the variation in the spectral bands.

A simple MLE classification for each pixel based on these PCs yielded an
accuracy of 66.9\%. Though 66.9\% might be considered low and significant
overlap of the classes appears to exist in the bottom panel of
Figure~\ref{fig:geopc}, this accuracy and overlap are expected to some degree
due to the nature of the data and ambiguity in some of the class definitions.
So, these results are pleasing.

\section{Conclusion}
The matrix normal distribution is a natural candidate for situations involving
some sort of structure or separability in the dimensions of the data. In this
article we derived an expectation-maximization algorithm for the matrix normal
distribution. A simulation study exploring different sample sizes and
proportions of missing data showed the usefulness and shortcomings of this
method when compared to a full, unconstrained multivariate normal
distribution. An example of this type of scenario, in the field of Remote
Sensing, produced physically useful and interpretable results. As data becomes
more abundant and higher in dimension the challenge of extracting information
continues to grow in difficulty and importance. The kronecker covariance
structure can provide both a richer physical interpretation of the parameters
as well as help the estimating procedure. Now, even with missing data,
accurate estimates of these parameters are obtainable.


\appendix
\section{EM Algorithm for Matrix Normal}

Here we give a more detailed implementation of the proposed EM method in
Algorithm~\ref{alg:em}. In what follows, the operator $\eop(\cdot)$ produces a
\emph{mask} matrix according to the corresponding row and column indices. Take
the following example:
\[
X_i = \left[ 
\begin{array}{ccccccc}
1 & 4 & 7 & 10 & ? & 16 & ? \\
? & 5 & 8 & 11 & 14 & 17 & 20 \\
? & 6 & ? & 12 & 15 & ? & 21
\end{array}
\right].
\]
There are six missing values that correspond to
$m_s(i) = [2 \, 3 \, 3 \, 1 \, 3 \, 1]$ and
$m_c(i) = [1 \, 1 \, 3 \, 5 \, 6 \, 7]$.
In this case, $\eop(m_c(i))$ is $6 \times 7$ (the number of missing
values by the number of columns of $X$) and $\eop(m_s(i))$ is
$6 \times 3$ (the number of missing values by the number of rows of $X$):
\[
\eop(m_c(i)) = \left[ 
\begin{array}{ccccccc}
1 & 0 & 0 & 0 & 0 & 0 & 0 \\
1 & 0 & 0 & 0 & 0 & 0 & 0 \\
0 & 0 & 1 & 0 & 0 & 0 & 0 \\
0 & 0 & 0 & 0 & 1 & 0 & 0 \\
0 & 0 & 0 & 0 & 0 & 1 & 0 \\
0 & 0 & 0 & 0 & 0 & 0 & 1
\end{array}
\right]
\quad\text{and}\quad
\eop(m_s(i)) = \left[ 
\begin{array}{ccc}
0 & 1 & 0 \\
0 & 0 & 1 \\
0 & 0 & 1 \\
1 & 0 & 0 \\
0 & 0 & 1 \\
1 & 0 & 0
\end{array}
\right].
\]
In this way, there is an ``indicator'' row for each index in $m_c(i)$ and
$m_s(i)$, respectively.

In Algorithm~\ref{alg:em} we denote by $\circ$ the Hadamard (element-wise)
product. We further denote by $A[m_s(i)]$ the rows of $A$ indexed by $m_s(i)$,
and by $A[-m_s(i)]$ the rows of $A$ there are \emph{not} in $m_s(i)$. A
similar notation is used to subset columns based on index sets.

\begin{algorithm*}[htbp]
\normalsize

\emph{Initialize}: $\mu^{(1)} \leftarrow 0_{p \times q}$;
$\Sigma_c^{(1)} \leftarrow I_q$;
$\Sigma_s^{(1)} \leftarrow I_p$\;

\For{$t \leftarrow 1, \ldots$ (until convergence)}{
  $\mu^{(t+1)} \leftarrow 0_{p \times q}$;
  $\Sigma_c^{(t+1)} \leftarrow 0_q$;
  $\Sigma_s^{(t+1)} \leftarrow 0_p$\;

  \For{$i \leftarrow 1, \ldots, n$}{
    \tcp{Initialize missing entry indices and conditional covariance:}
    \emph{Set} {\sf miss}$(i)$ as the indices of missing entries in the $i$-th
    observation\;

    \emph{Set} $m_s(i)$ and $m_c(i)$ as the row and column indices of the
    entries in {\sf miss}$(i)$\;

    \emph{SWEEP} the {\sf miss}$(i)$ rows of
    $R = \Sigma_c^{-1^{(t)}} \otimes \Sigma_s^{-1^{(t)}}$
    \tcp*{Compute $\var_{Z_i \given Y_i}[X_i]$}

    \BlankLine\BlankLine

    \tcp{Set auxiliary variables:}
    $X_m \leftarrow \left[\begin{array}{c}
    Y_i \\
    \mu^{(t)}[\text{\sf miss}(i)] +
    R[-\text{\sf miss}(i), \text{\sf miss}(i)]^\top
    (Y_i - \mu^{(t)}[-\text{\sf miss}(i)])
    \end{array} \right]$
    \tcp*{$\Exp_{Z_i \given Y_i}[X_i]$}

    $R_m \leftarrow R[\text{\sf miss}(i), \text{\sf miss}(i)]$\;

    $S_{s,m} \leftarrow \Sigma_s^{-1^{(t)}}[m_s(i), m_s(i)]$\;

    $S_{c,m} \leftarrow \Sigma_c^{-1^{(t)}}[m_c(i), m_c(i)]$\;

    \BlankLine\BlankLine

    \tcp{Update:}
    $\mu^{(t+1)} \leftarrow \mu^{(t+1)} + X_m$\;

    $\Sigma_c^{-1^{(t+1)}} \leftarrow \Sigma_c^{-1^{(t+1)}} +
    \eop(m_c(i))^\top \cdot (R_m \, \circ \, S_{c,m}) \cdot \eop(m_c(i)) +
    (X_m - \mu^{(t)})^\top \Sigma_s^{-1^{(t)}} (X_m - \mu^{(t)})$\;

    $\Sigma_s^{-1^{(t+1)}} \leftarrow \Sigma_s^{-1^{(t+1)}} +
    \eop(m_s(i))^\top \cdot (R_m \, \circ \, S_{s,m}) \cdot \eop(m_s(i)) +
    (X_m - \mu^{(t)}) \Sigma_c^{-1^{(t)}} (X_m - \mu^{(t)})^\top$\;

    $\sigma^{2^{(t+1)}} \leftarrow \sigma^{2^{(t+1)}} +
    \sum_j \V{(R_m \, \circ \, S_{c,m} \, \circ \, S_{s,m})} +
    \tr(\V{(X_m - \mu^{(t)})^\top}
    (\Sigma_c^{-1^{(t)}} \otimes \Sigma_s^{-1^{(t)}})
    \V{(X_m - \mu^{(t)})})$\;
  }

  \BlankLine\BlankLine

  \tcp{Scale:}
  $\mu^{(t+1)} \leftarrow \mu^{(t+1)}/n$\;
  $\Sigma_c^{(t+1)} \leftarrow \Sigma_c^{(t+1)}/\Sigma_{c,11}^{(t+1)}$\;
  $\Sigma_s^{(t+1)} \leftarrow \Sigma_s^{(t+1)}/\Sigma_{s,11}^{(t+1)}$\;
  $\sigma^{2^{(t+1)}} \leftarrow \sigma^{2^{(t+1)}}/(nBT)$\;
}
\caption{Matrix Normal Expectation-Maximization} 
\label{alg:em}
\end{algorithm*}

\bibliographystyle{spbasic} 
\bibliography{mn-em}

\end{document}